# Graphene at high bias: cracking, layer by layer sublimation and fusing

*A. Barreiro[1,†,*], F. Börrnert[2,†], M. H. Rümmeli[2,3], B. Büchner[2], L. M. K. Vandersypen[1]*

[1] Kavli Institute of Nanoscience, Delft University of Technology, Lorentzweg 1, 2628 CJ Delft, The Netherlands.

[2] IFW Dresden, Postfach 270116, 01171 Dresden, Germany

[3] TU Dresden, 01069 Dresden, Germany

[†] these authors contributed equally

* ab3690@columbia.edu

ABSTRACT. Graphene and few-layer graphene at high bias expose a wealth of phenomena due to the high temperatures reached. With in-situ transmission electron microscopy (TEM) we observe directly how the current modifies the structure, and vice versa. In some samples, cracks propagate from the edges of the flakes, leading to the formation of narrow constrictions or to nanometer spaced gaps after breakdown. In other samples we find layer-by-layer evaporation of few-layer graphene, which could be exploited for the controlled production of single layer graphene from multi-layered samples. Surprisingly, we even find that two pieces of graphene that overlap can heal out at high bias and form one continuous sheet. These findings open up new avenues to structure graphene for specific device applications.

KEYWORDS. Graphene, transmission electron microscopy, nanostructuring, nano-sculpting, in-situ Joule heating.

MANUSCRIPT TEXT. Since the first isolation of graphene in 2005 this material has attracted intense interest for a wide range of electronics applications.[1] Novel devices such as field-effect transistors (FETs) based on nanoribbons,[2,3] optoelectronics devices with monolayer-bilayer junctions,[4,5] and nanometer spaced electrodes for molecular junctions,[6] require very specific nano-engineering techniques for patterning and structuring the graphene. Conventional lithography is in many cases not sufficient and in-situ techniques such as current-induced annealing have proven very useful.[7] For instance, by applying a high bias, the mobility of graphene can be significantly improved,[7,8] narrow constrictions that behave as quantum point contacts can be formed,[9] and nanometer spaced gaps that are stable at room temperature can be controllably formed.[6]

In this Letter, we perform real-time in-situ TEM studies of graphene at high bias. We report a rich variety of phenomena that provide important insights into how to shape graphene or modify its structure (e.g. number of layers) by Joule heating. We observe peeling off of multilayered suspended graphene sheets layer-by-layer locally until only a graphene monolayer remains. Moreover, we are able to controllably narrow down graphene into nanoribbons as narrow as 1 nm, which sustain current densities as high as $6 \cdot 10^9$ Acm$^{-2}$, in agreement with a recent report by Lu et al.[10] Surprisingly, we also find that the breakdown current density sharply increases with decreasing width. Finally, two separate but overlapping pieces of graphene can become one continuous sheet again. The results offer a new approach to structuring graphene that is relevant for specific device applications.

Chips with single-layer and few-layer graphene flakes supported by metal contacts were mounted on a custom-built sample holder for TEM with electric terminals, enabling simultaneous TEM imaging and electrical measurements. For imaging, a FEI Titan$^3$ 80–300 transmission electron microscope with a CEOS third-order spherical aberration corrector for the objective lens was used. It operated at an acceleration voltage of 80 kV to reduce knock-on damage. All studies were conducted at room temperature with a pressure of approx. $10^{-7}$ mbar. Figure 1 shows an image of an electrically contacted few-layer graphene device inside the TEM. The sample fabrication procedure is described in detail in the supporting information. In total, we measured 15 devices.

First, we perform in-situ current-induced annealing of the suspended graphene devices by taking the samples to the high bias regime, specifically up to 2 - 3 V.[7] Temperatures as high as 2000 °C are reached due to Joule heating.[11,12] As a result, most contaminants from fabrication are removed and we observe in the TEM that we obtain atomically clean graphene devices.

If we increase the bias even further, we reach the high-current limit. In this regime, the samples are at such a high bias that they are close to a complete and irreversible electrical breakdown. Because of this, we increase the bias very carefully in steps of 10 mV until we see that the current flowing through the sample decreases as a function of time, and then keep the bias constant (typically around 3 V). At this constant bias we observe that the total current flowing through the sample further decreases over time, which is an indication of carbon atom sublimation.[11] The most frequent situation is that a crack forms on one edge of the sample half-way between the electrodes and slowly propagates towards the other edge of the device. This can be understood by the fact that removing an atom from a vacancy edge requires much less energy (~7 eV) than that from a perfect lattice site (~30 eV).[13] In few-layer graphene samples, the cracks in the different layers are closely spaced and propagate in the same direction with a similar speed. When the further advanced crack reaches the other side of the sample,[14] it changes direction and moves towards the other crack until the two cuts meet and the sample ends up with two separate but very closely spaced sheets (Fig. 2 and video S1).[11] A similar mechanism has been reported for the formation of nanometer spaced gaps in mono- and few layer graphene on $SiO_2$.[6,14] As explained in detail in section S3 of the supporting information, the main driving mechanism of carbon atom sublimation in our experiments should originate from Joule heating.

Importantly, the propagation of cracks can be harnessed to form very narrow graphene nanoconstrictions (GNCs) and can be applied to the formation of nanoribbons. For this purpose, the crack propagation must be controllably stopped before complete breakdown, see figures 3 and 4, and video S2. In these two specific cases edges exhibiting a strong contrast can be observed, suggesting a bilayer edge (BLE), in contrast to a faint contrast, indicating a monolayer edge (MLE).[14,16-18] In the measurements corresponding to figure 4, we observed a stepwise decrease in the current as the constriction was narrowed. From real-time imaging in the TEM, we could infer that these steps corresponded to structural changes in the constriction. Interestingly, the device in figure 4 originated from merging two separated graphene layers, see figure S5.[14]

Remarkably, these nanometer sized constrictions are able to hold together the bigger parts of the flake that are connected to the electrodes and exhibit a defect-free lattice as resolved by aberration corrected high resolution TEM (AC-HRTEM) in figure 3.

Moreover, the GNCs are also able to sustain enormous current densities before breakdown ($j_{BR}$).

Indeed, $j_{BR}$=40 μA/nm, corresponding to $6 \cdot 10^9$ Acm$^{-2}$ if normalized for a graphene thickness of 0.68 nm for the 1 nm wide constriction in figure 4c, can be extracted from the I-V data in figure 4e. Recently, a slightly higher $j_{BR}$ has been observed for GNCs sculpted in-situ with the TEM beam.[10] For comparison, the $j_{BR}$ of carbon nanotubes (CNTs) corresponds to a current density exceeding $10^9$ Acm$^{-2}$,[19,20] or even up to $4.5 \cdot 10^9$ Acm$^{-2}$ for very short (in the 50 nm range) single-wall CNTs,[21,22] comparable to the $j_{BR}$ that our graphene nanoribbons are able to sustain.

For both CNTs and graphene nanoribbons, $j_{BR}$ is several orders of magnitude larger than in present-day interconnects.[23] It is also around 2 orders of magnitude larger than the values reported for 200 nm wide suspended graphene constrictions,[24] and for μm sized few-layer graphene samples.[11,17] Consistent with those reports, we observe that the breakdown current density $j_{BR}$ sharply decreases with increasing width of the graphene device, down to only $1.2 \cdot 10^7$ Acm$^{-2}$ for a 800 nm wide piece (see figure 5). Indeed, it was not possible to narrow down all the devices to nanoconstrictions. Here the width of the flake refers to the width of the device just before complete electrical breakdown.

It is surprising that the suspended GNCs are able to sustain a $j_{BR}$ more than 2 orders of magnitude bigger than a μm wide suspended graphene. Based on the information we have it is difficult to be certain about the origin of this observation. Although we do observe that we have rather clean edges in the GNCs, we don't believe the edges play a role in enhancing $j_{BR}$ as they introduce an additional scattering source as compared to the bulk, which should result in smaller breakdown current densities for narrower ribbons, opposite to what we observe. One possibility why $j_{BR}$ increases for narrower constrictions is a more efficient heat dissipation of the short nanoribbons through the much wider graphene counterparts that connect them to the metal electrodes. Also it could simply be the case that graphene flakes with more adsorbates break down earlier in an uncontrollable manner due to a sudden reaction with the contaminants, leading to a complete electrical breakdown. Cleaner samples allow for a controlled narrowing leading to GNCs, therefore sustaining higher $j_{BR}$ as the breakdown is not triggered by contaminants.

On some occasions, we find that carbon atom sublimation occurs not only in the form of cracks starting from the edges but also in the central area of the flake, eventually leading to layer-by-layer sublimation. For example, during the crack propagation of two BLEs marked by arrows in figure 6 (a) towards the central region of the flake we found that suddenly one of the two layers developed a hole in the center of the constriction and propagated outwards in a polygonal fashion, before the remaining layer(s) eventually broke down (fig. 6 and video S3). The lighter contrast in the broken region and the fact that there were no more lattice fringes at the sides of the flake suggest that a monolayer was present just before breakdown. This finding has been observed not only for bilayer graphene but also for multilayered samples where layer-by-layer sublimation eventually leads to the formation of single layer graphene (see fig. 7 and video S4).

Possible reasons for preferential carbon atom sublimation starting from the center rather than by continuing the crack propagation could be higher temperatures reached in the middle of the flake or the presence of a defect in the lattice from where the atom sublimation ignites. In total we have observed a similar behaviour in 3 samples while steady crack propagation until the final breakdown was observed in 8 devices, illustrating the various types of behavior that occur at high bias, close to electrical breakdown. When sufficiently well understood, controlled sublimation may be used for tailoring layer thickness, e.g. by creating damage on purpose in the center and next applying a large bias. On the other hand, sublimation from the center could present a problem for controlled crack propagation and the formation of narrow constrictions if unintentional defects exist in the middle of the flake.

Another interesting event we found when applying a high bias voltage is that two pieces of graphene resulting from rupture of a flake can overlap and start conducting again, see figures S5-S8.[14] In figure 8 a,b the region of overlap can be identified by its darker contrast and by following the dotted lines that are guides to the eye indicating the edges of the respective pieces formed upon rupture. The area of overlap changes over time (Figure 8a,b and videos S5 and S6),[14] and these changes are accompanied by conductance changes of the device, see first arrow in figure 8g and figure S6.[14] In figure S10 several conductance changes can be observed, corresponding to the repeated motion of one flake relative to the other, seen in video S6. We have observed overlapping flakes after rupture with changes in conductance on 4 devices.

Perhaps the most surprising finding is that the overlapping regions can heal to form one continuous, clean graphene layer, see figure 8c (the graphene patches on top that originated from amorphous carbon have sublimated in the process)[10,25]. The overlap area is very hot as it is located in the central part of the flake and has the highest resistance as there electron transport occurs from flake to flake. As a result of the high temperature, the graphene heals out into a seamless graphene sheet. From the dark contrast of the edges in figure 8 d,e we infer that this is, again, bilayer graphene.

At the moment when the graphene grows together from the two overlapping regions, a sudden increase in conductance is observed, despite the simultaneous reduction in width (second arrow in figure 8f); this can be expected as the resistance through a seamless graphene sheet is smaller than through two overlapping sheets where the electrons have to hop from one sheet to the other. We note that we have never observed a sudden increase of current upon evaporating graphene; it is the healing which causes the increased conductance. While keeping the bias constant, the newly formed seamless graphene next narrows down gradually by crack propagation from the edges until a constriction of only a few nanometers is formed, see figure 8 d,e.

In conclusion, carbon atom sublimation driven by a high bias can represent a versatile and efficient alternative to beam-driven erosion of carbon atoms for nanostructuring graphene. *Via* in-situ TEM studies in the high current limit, we observe real-time formation of cracks that lead to ultra-narrow constrictions, layer-by-layer removal, and the mechanical motion of two disconnected graphene layers one on top of the other that can heal into a perfect defect-free graphene. A more detailed understanding of the dynamics of layer-by-layer peeling and narrowing of few-layer flakes, may provide tools for tailoring the graphene layer thickness and lateral dimensions with atomic precision, enabling new device applications. When sufficiently well understood and controlled, this technique could be applied without the visual feedback from in-situ TEM measurements, so that it doesn't rely on expensive equipment.

ACKNOWLEDGMENT. We gratefully acknowledge M. Rudneva and H. Zandbergen for help in the early stages of the experiment, G. F. Schneider for help with graphene transfer and M. Zuiddam for help with the deep reactive ion etching process. Financial support was obtained from the Dutch Foundation for Fundamental Research on Matter (FOM), AGAUR (2010_BP_A_00301), DFG (RU1540/8-1), EU (ECEMP) and the Freistaat Sachsen.

SUPPORTING INFORMATION PARAGRAPH. **Supporting Information Available.** Details of the sample fabrication procedure and the experimental methods; effect of the 80 keV TEM electron beam; additional TEM images of the crack formation leading to a narrow constriction and the conductance switches due to the changes in the area of two overlapping graphene flakes; and MWNT formation at the edges of few-layer graphene after electrical breakdown.

Movie S1: Video of the crack propagation in figure 2, thirteen times faster than in real-time.

Movie S2: Video of the formation of the narrow constriction at high bias in figure 3, four times faster than in real-time.

Movie S3: Video of the sublimation of one of the layers of a bilayer in figure 6, two times faster than in real-time.

Movie S4: Video of the layer-by-layer sublimation in figure 7, six times faster than in real-time.

Movie S5: Video of the mechanical switches of two overlapping graphene layers and transformation into a seamless graphene sheet in figure 8 in real-time.

Movie S6: Video of the two overlapping layers in figure 8 sliding on top of each other a large number of times, 32 times faster than in real-time.

FIGURE CAPTIONS.

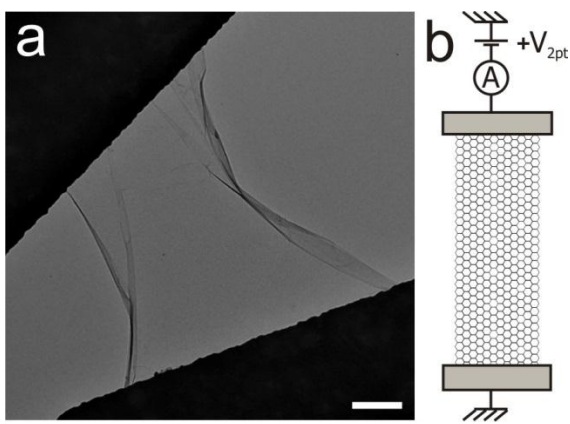

**Figure 1.** (a) TEM image of an as-fabricated few-layer graphene flake connected to two Cr/Au electrodes. The scale bar is 200 nm. (b) Schematic representation of the measurement set-up. The device is voltage biased and the current is measured.

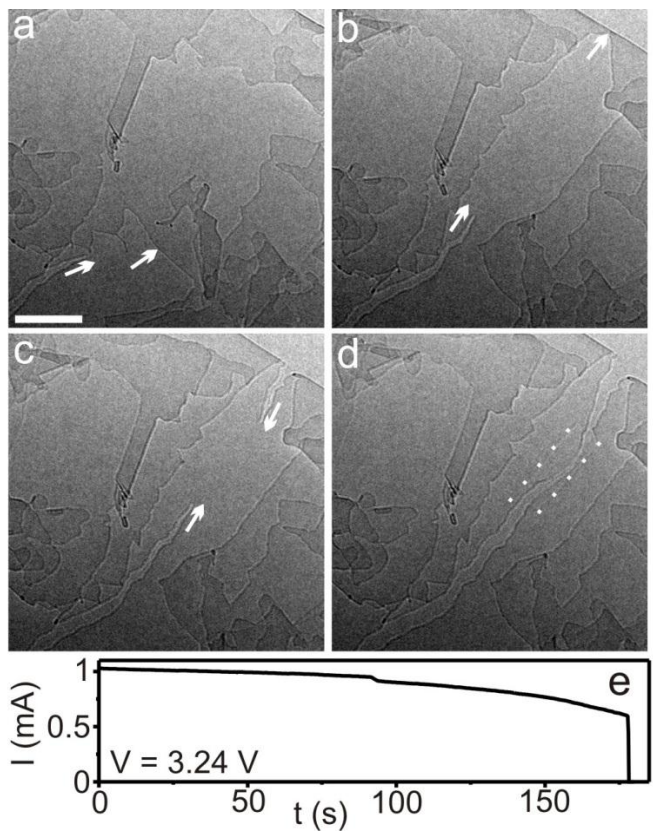

**Figure 2.** Evolution of a crack. The arrows point in the direction of propagation of the crack. (a) The crack propagates in two different bi-layers. The scale bar is 100 nm. (b) The top bi-layer crack reaches the edge of the sample and (c) reverses its direction of motion.[14] (d) The two wedges propagate towards each other until they meet. A graphene monolayer region (marked by dotted lines) forms a nanometer spaced gap. The time elapsed between the 4 frames is 100 s (Movie S1 shows the entire process). (e) The current (I) flowing through the device as a function of time decreases steadily during the electro-burning while keeping the bias voltage fixed at 3.24 V. The final breakdown current density is $4.68 \cdot 10^8$ $Acm^{-2}$ when normalized to the thickness of a bilayer graphene (0.68 nm).

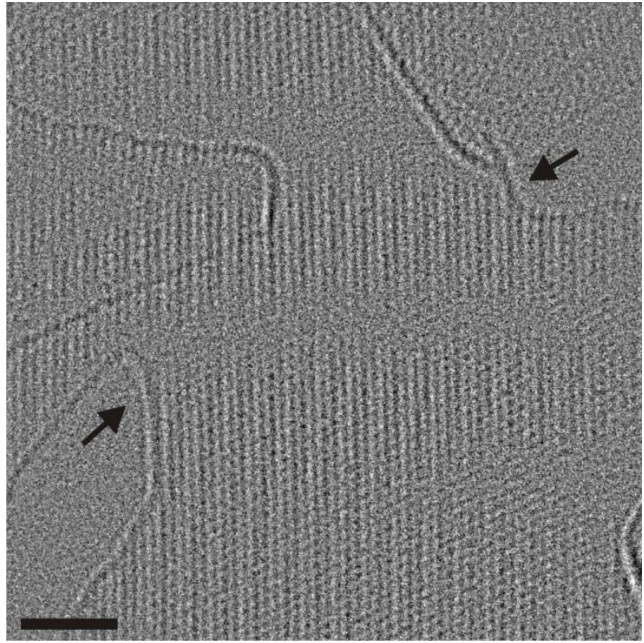

**Figure 3.** AC-HRTEM micrograph of a narrow bilayer graphene constriction with a very regular and defect-free lattice, formed by controllable electro-burning by crack propagation from the sides. The scale bar is 2 nm. This micrograph has been subjected to a Wiener filter to remove the background and enhance the signal to noise ratio.

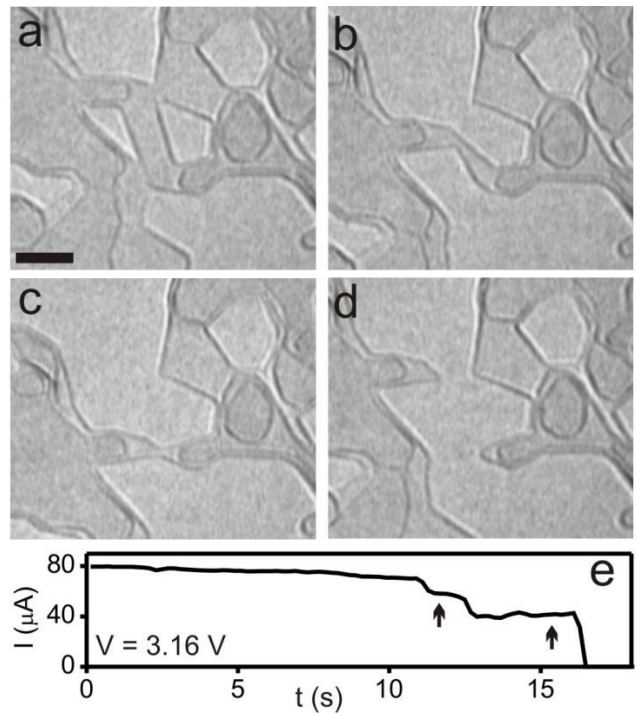

**Figure 4.** Evolution of a narrow bilayer graphene constriction at high bias. The graphene patches on the bigger parts of the flake grew from amorphous carbon deposited on the sample at zero bias.[12,25] (a) Two narrow constrictions are connected to two big pieces of graphene. The scale bar is 5 nm. (b) Rupture of one of the junctions, leading to a single narrow constriction with a kink. (c) Removal of the kink and gradual narrowing of the constriction. (d) Rupture of the narrow constriction.. The time elapsed between the 4 frames is 18 s. (e) Current (I) as a function of time at a fixed bias voltage V = 3.16 V of panels (a) – (d). The arrows correspond to panel (b) and (c), respectively. The final breakdown current density $j_{BR}$ is $6 \cdot 10^9$ Acm$^{-2}$ when normalized to the thickness of bilayer graphene (0.68 nm).

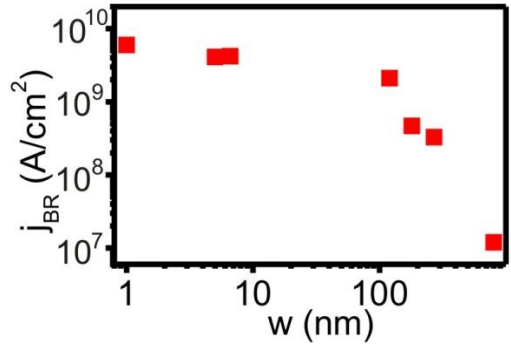

**Figure 5.** Breakdown current density $j_{BR}$ as a function of width in the range of 1 – 800 nm.

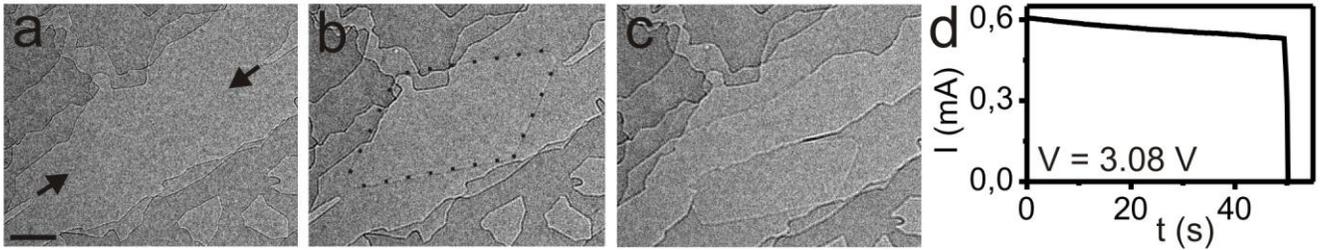

**Figure 6.** (a) Bilayer graphene sample which was narrowed down by crack propagation from both sides. The arrows point in the direction of crack propagation from both sides. The scale bar is 20 nm. (b) The cracks at the sides stop propagating and a hole forms in the middle of the top layer which expands in a polygonal fashion leaving a monolayer of graphene. The dotted lines are guides to the eye. (c) Electrical breakdown of the device. The time elapsed between the 3 frames is 50 s. (d) Current as a function of time at V=3.08V. The breakdown current density $j_{BR}$ is $2.1 \cdot 10^9$ Acm$^{-2}$.

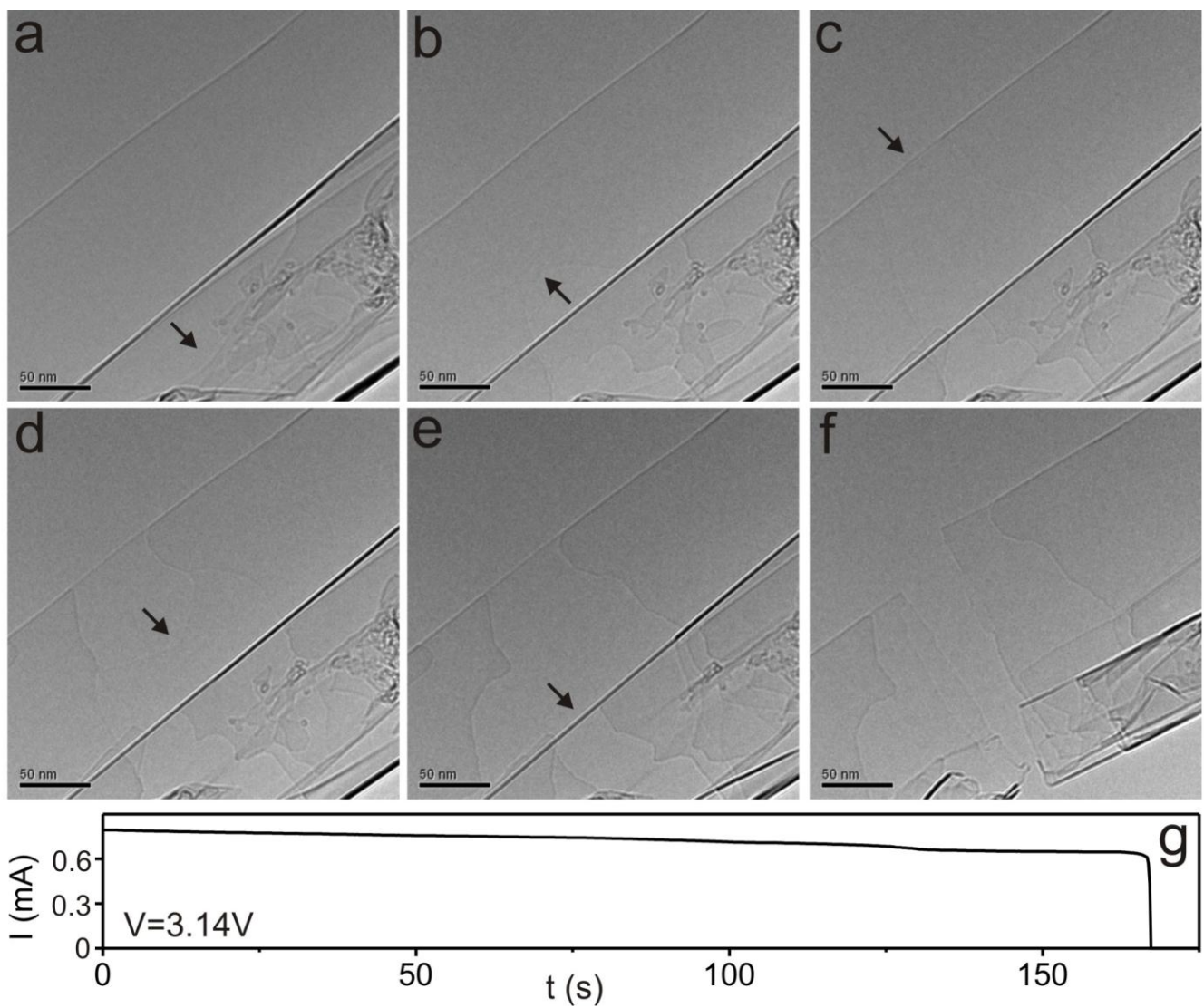

**Figure 7.** (a-e) Layer-by layer sublimation of a folded few-layer graphene flake until complete electrical breakdown (f). The arrows point in the direction of propagation of the layer sublimation. The time elapsed between the 6 frames is 140 s. (g) Current vs time at V=3.14V. The breakdown current density $j_{BR}$ is $3.3 \cdot 10^8$ Acm$^{-2}$.

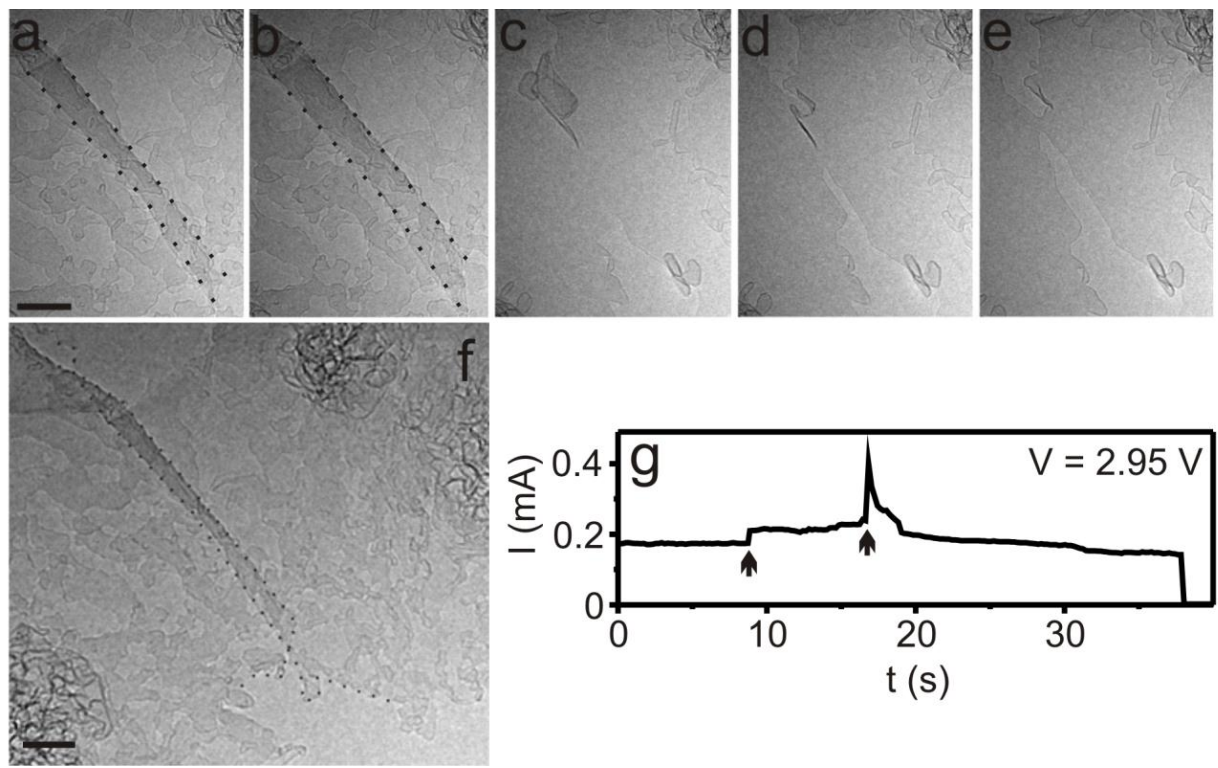

**Figure 8.** (a,b) TEM images of two overlapping regions of graphene which move as a function of time. The dotted lines are guides to the eye indicating the borders of the two graphene layers. They enclose the region where the two layers overlap (darker contrast in the TEM image) and flake-to-flake electron transport occurs. The scale bar is 20 nm. (c) Sudden healing of the two overlapping flakes and fusing into continuous, seamless bilayer graphene with a smallest width of approximately 60 nm. The constriction next narrows down gradually by atom sublimation from the edges due to Joule heating, reaching a width of (d) 20 and (e) 5 nm, respectively. The time elapsed between the 5 frames is 35 s (see also Video S5). (f) Zoomed out TEM image of figure 8a. The scale bar is 20 nm. For a further magnified image please see figure S9.[14] (g) Current as a function of time at V=2.95V. The arrows correspond to panels (b) and (c), respectively. The breakdown current density $j_{BR}$ is $8.1 \cdot 10^9$ Acm$^{-2}$.

# Graphene at high bias:

# cracking, layer by layer sublimation and fusing

# (Supporting Information)

*A. Barreiro*[1,†,*], *F. Börrnert*[2,†], *M. H. Rümmeli*[2,3], *B. Büchner*[2], *L. M. K. Vandersypen*[1]

[1] Kavli Institute of Nanoscience, Delft University of Technology, Lorentzweg 1, 2628 CJ Delft, The Netherlands.

[2] IFW Dresden, Postfach 270116, 01171 Dresden, Germany

[3] TU Dresden, 01069 Dresden, Germany

[†] these authors contributed equally

\* ab3690@columbia.edu

### Section 1. Sample fabrication

We here describe the different steps for our sample fabrication procedure, see Fig. S1. Double side polished wafers with 300 μm thickness of Si and 1.3 μm thermal $SiO_2$ on both sides are used. A multilayer etch-mask is fabricated on the backside of the wafer to be able to etch through the 300 μm of Si: further 4.2 μm $SiO_2$ are added to the thermal $SiO_2$ and 540 nm Si is deposited on top of the $SiO_2$. Both layers are grown by plasma-enhanced (PE) chemical vapor deposition (CVD). Then a thick ZEP resist mask is spun on top and used for patterning 50 x 50 μm squares by e-beam lithography. Now we etch through the multilayer etch-mask in steps. First, the 540 nm of Si in the holes patterned by e-beam are etched away by a highly anisotropic deep reactive ion etching (RIE) step, commonly known as Bosch etching [1] and with which it is possible to create very steep and almost perpendicularly etched walls. Then the Si acts as a mask to etch the holes further through the entire $SiO_2$ by a dry RIE step. Now the 5.5 μm of $SiO_2$ (1.3 μm thermal $SiO_2$ plus 4.2 μm PECVD $SiO_2$) act as a mask to etch through the entire 300 μm of Si by a further Bosch etching step. As a result, on the front side of the wafer freestanding $SiO_2$ membranes are left. Electrodes are patterned on the $SiO_2$ membranes using electron-beam lithography and subsequent evaporation of Cr/Au (10 nm / 150 nm). Finally the $SiO_2$ membranes are completely removed by means of a HF etching step. As a result, the thick electrodes remain free-standing without any substrate. The next step is to transfer graphene flakes on top of the free-standing electrodes.

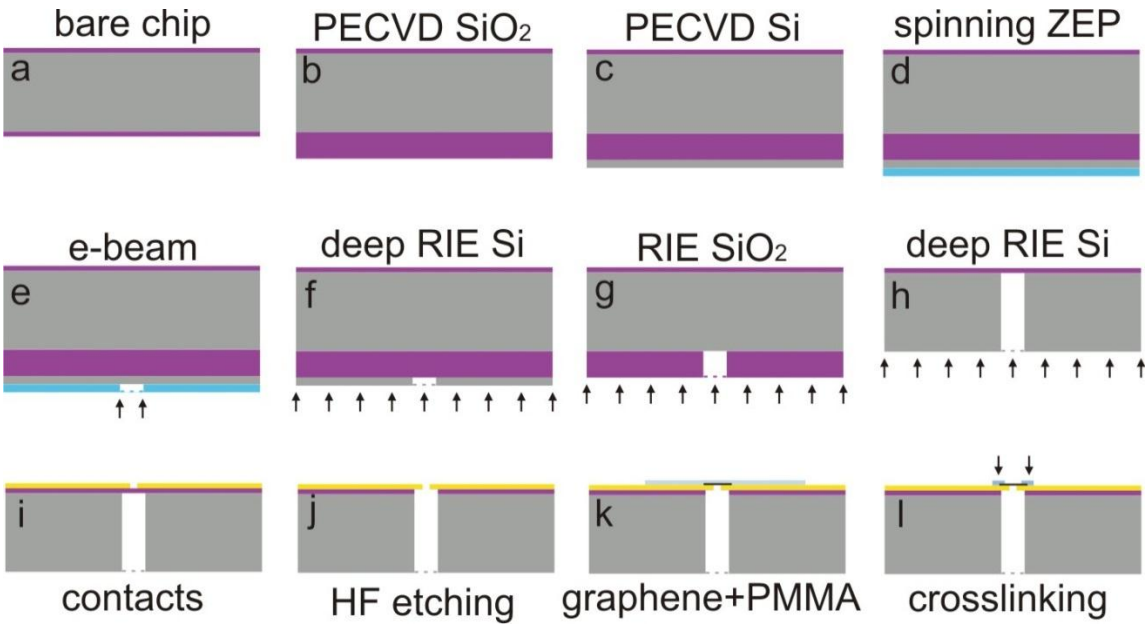

**Figure S1**. Schematic illustration of the different steps during the sample fabrication procedure. (a) Double side polished wafers with thermal $SiO_2$ on both sides are used. (b, c, d) A multilayer etch-mask is fabricated on the backside of the wafer to be able to etch through the whole Si: (b) further 4.2 μm

SiO$_2$ are added to the thermal SiO$_2$ and (c) 540 nm Si is deposited on top of the SiO$_2$. Both layers are grown by PE-CVD. (d) A thick ZEP resist mask is spun on top and (e) used for patterning 50 x 50 µm squares by e-beam lithography. (f, g, h) Stepwise etching through the multilayer etchmask. (f) First, the 540 nm of Si in the holes patterned by e-beam are etched away by a "Bosch" deep reactive ion etching (RIE) step [1]. (g) Si acts as a mask to etch the holes further through the entire SiO$_2$ by a dry RIE step. (h) SiO$_2$ acts as a mask to etch through the entire 300 µm of Si by a further Bosch etching step. As a result, on the front side of the wafer freestanding SiO$_2$ membranes are left. (i) Electrodes are patterned on the SiO$_2$ membranes using electron-beam lithography and subsequent evaporation of Cr/Au (10 nm / 150 nm). (j) The SiO$_2$ membranes are completely removed by means of a HF etching step. The thick electrodes remain free-standing without any substrate. (k) Alignment of a graphene embedded in a PMMA film on top of the free-standing electrodes. (l) Fabrication of clamps from crosslinked PMMA to prevent that the graphene rolls up during the subsequent lift-off in acetone and boiling IPA.

For this purpose, graphene flakes are obtained by mechanical exfoliation of Kish graphite on silicon wafers coated with a 280 nm thick silicon oxide layer. Few layer graphene flakes are identified using optical microscopy. PMMA is spun on the wafers and the whole wafer is left floating in a 1M NaOH solution over night. The silicon dioxide layer is slowly etched by the NaOH and the NaOH intercalates between the PMMA and the wafer. As a result, the wafer falls off while the PMMA film with the graphene flakes attached to it is released and floats on top of the NaOH solution [2]. The PMMA film is then transferred without drying into a Petri dish with water and the graphene flake attached to the PMMA film is aligned onto the electrodes [3]. Afterwards, the regions of the PMMA film where the graphene and the Au electrodes overlap are strongly overexposed by means of e-beam lithography so that the PMMA crosslinks. As a result, clamps are formed from the crosslinked PMMA that firmly fix the graphene onto the electrodes and hinder it from rolling up during the subsequent lift-off in acetone and boiling IPA. Figure S2 shows a SEM image of a device at the end of the fabrication process.

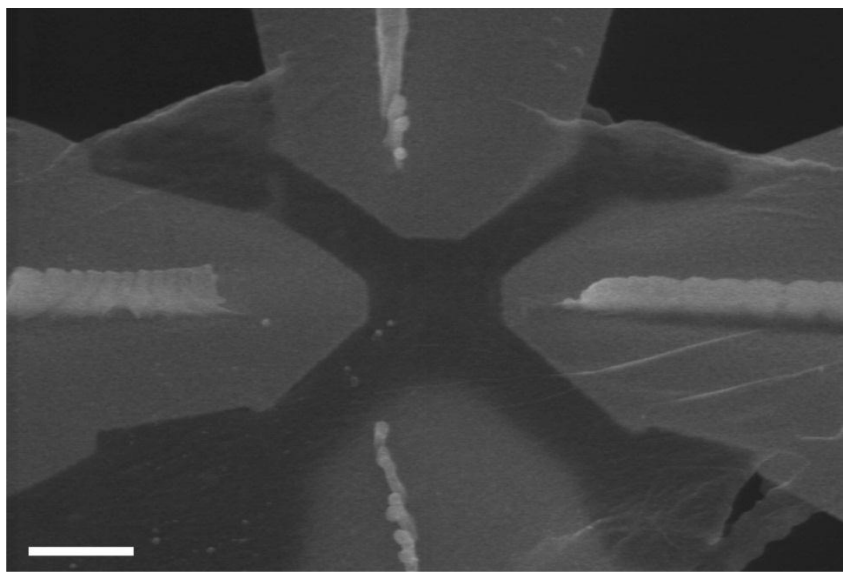

**Figure S2.** SEM image of a few-layer graphene flake aligned on top of 4 Cr/Au electrodes and clamped by cross-linked PMMA. The scale bar is 1 μm.

**Section 2. Experimental methods**

For imaging, a FEI Titan$^3$ 80–300 transmission electron microscope with a CEOS third-order spherical aberration corrector for the objective lens was used. It was operated at an acceleration voltage of 80 kV to reduce knock-on damage. The images were recorded with a Gatan UltraScan 1000 camera via the Gatan DigitalMicrograph software. To enhance the temporal resolution for in situ observation down to 350 ms per frame the camera was used in conjunction with the TechSmith Camtasia Studio screen recorder software at 4 pixel binning with an acquisition time of 0.05 s. All studies were conducted at room temperature and a pressure around $10^{-7}$ mbar.

**Section 3. Effect of the 80 keV electron beam**

The atom sublimation that assists the phenomena we observe in the high-current limit is induced mainly by Joule heating, possibly assisted by the high-energy electron beam. Importantly, with the electron beam present but without Joule heating, neither crack propagation nor layer-by-layer peeling occurred. Instead, we observed amorphous carbon deposition [4, 5], and, at certain positions, holes created due to beam-driven chemical modifications below the knock-on damage threshold.

All the experiments described in this Letter were imaged with an 80 keV beam. It has been shown that 80 keV electron irradiation does not produce defects in a clean lattice [6]. However, current annealing in the high current limit can lead to temperatures as high as 2000 °C [7, 8]. Such high temperatures can not be reached only with the TEM beam. During TEM imaging without concomitant current annealing we find two different effects depending on the cleanliness of the graphene. On the one hand, imaging of clean current annealed samples leads to the deposition of increasing quantities of amorphous carbon [4, 5]. On the other hand, imaging of graphene flakes without a previous current annealing step, exposes fabrication residues to the electron beam. The reaction of the contaminants with the graphene by beam-driven chemical modifications at energies below the knock-on threshold [9] leads to the rupture of the flakes.

Therefore, the main driving mechanism of carbon atom sublimation should originate from Joule heating, although we cannot exclude that there might be a contribution to the carbon atom sublimation or the healing of graphene from the electron beam, too.

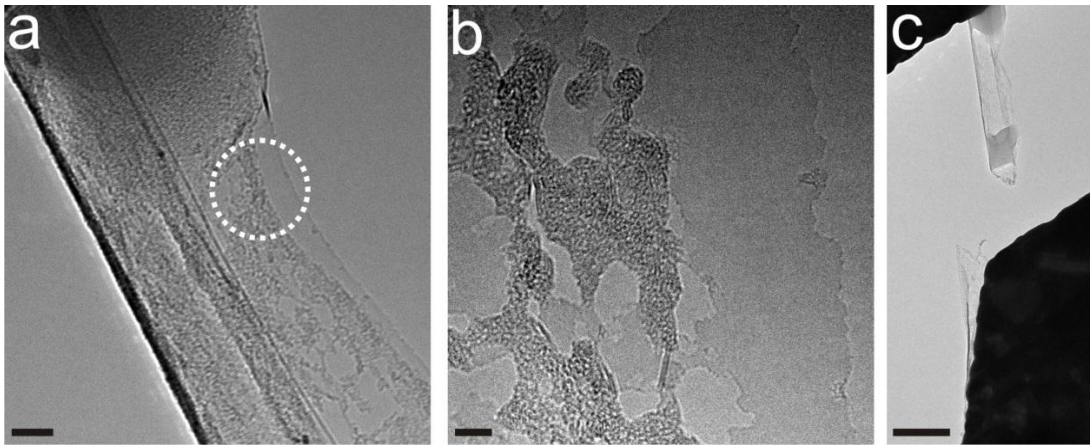

**Figure S3.** (a) TEM image of a few-layer graphene flake with fabrication residues. The scale bar is 20 nm. (b) Hole formation due to beam-driven chemical modifications in the lattice with the contaminants. The scale bar is 5 nm. (c) Rupture of the graphene due to an excessive accumulation of holes. The scale bar is 200 nm.

**Section 4. Crack evolution to form narrow constrictions**

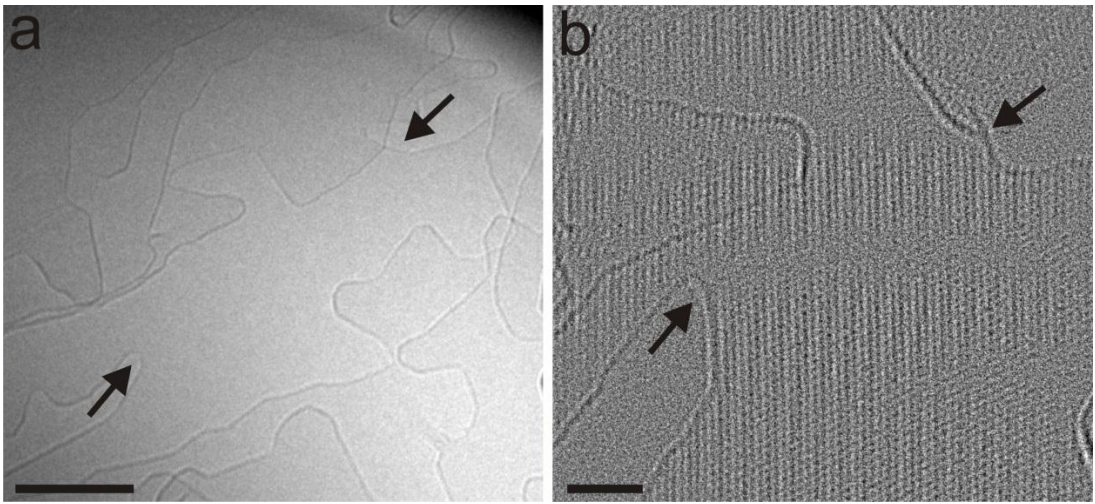

**Figure S4.** (a) Evolution of crack propagation from both edges in the final stages. The scale bar is 10 nm. (b) Narrow constriction with a very regular and defect-free lattice, formed by controllable electro-burning by crack propagation from the sides. The scale bar is 2 nm.

**Section 5. Discussion regarding the edges in figure 2 of the manuscript**

References 13 and 14 report that edges exhibiting a strong contrast correspond to BLEs, while a faint contrast of an edge indicates a MLE. Interestingly, from the dark contrast of the edges in figures 2 b,c it can be inferred that the edges of the two layers seem to be bonded together, possibly due to the high bias treatment [10]. At high temperatures, the edges of two graphene layers stacked on top of each other merge by a nanoarch to form bilayer edges (BLE) [10-13]. BLEs are much more stable than monolayer edges (MLE) because of the absence of dangling bonds.

Even multiple nested BLEs can form.[14] Indeed, in figure 2 d at the final stage of the breakdown it can be observed that an additional faint edge appears which is marked by white dotted lines, suggesting that the propagation of the cracks was happening in two nested bilayer graphene sheets, consisting of 4 graphene layers in total. This observation illustrates that determining the number of layers just by counting lattice fringes at graphene edges is problematic: a single fringe can represent either a MLE or a BLE [12]. A BLE usually shows much darker contrast compared to a MLE under the same imaging conditions, while MLEs show a faint contrast.

**Section 6. Merging of broken graphene layers and additional TEM images of the device in figure 4**

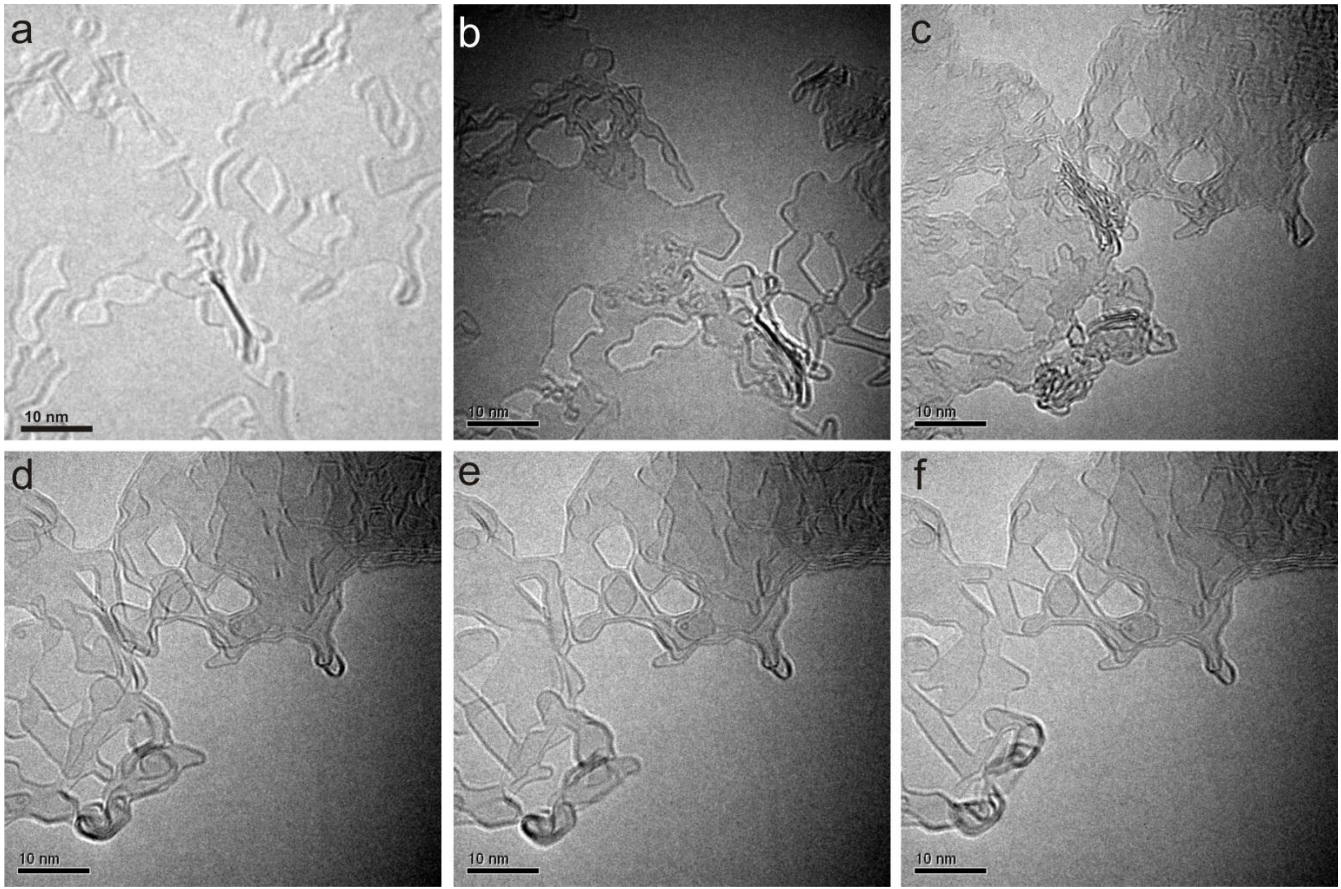

**Figure S5.** (a) Two separated graphene flakes that are not electrically connected. The scale bar is 10 nm. (b) Merging of the two separated flakes at zero bias. (c) Deposition of amorphous carbon due to TEM imaging at zero bias. (d-e) Carbon atom sublimation and gradual healing into a crystalline graphene lattice by current-induced annealing [5, 8].

## Section 7. Overlaps of broken graphene layers and conductance switches

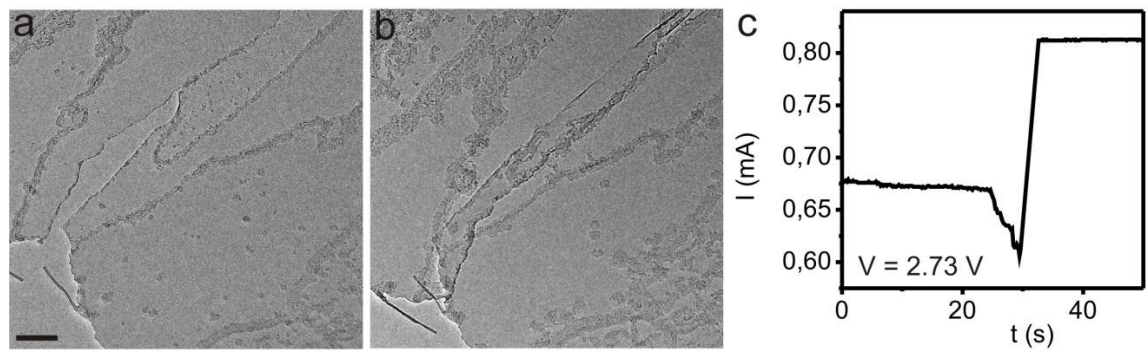

**Figure S6.** (a) Partly cracked graphene. Note that a MWNT forms at the edges of the flake after the first electro-burning events (see also fig. S8). The scale bar is 50 nm. (b) Overlapping of the cracked part at a high bias voltage of 2.73 V, giving rise to a switch in conductance. (c) Current vs. time at a constant bias voltage of 2.73 V. The sudden increase in current corresponds to the sudden overlapping. and touching of the two parts.

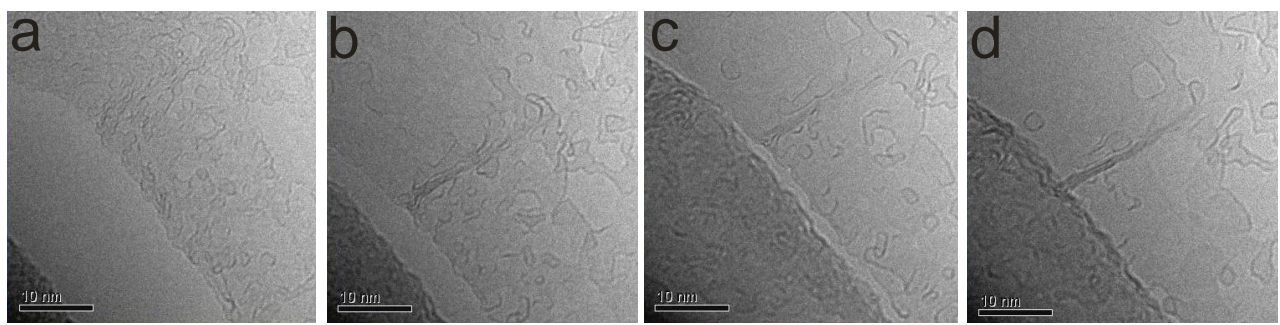

**Figure S7.** (a-c) TEM images of the stepwise approaching of two parts of a flake, (d) until finally forming an overlap.

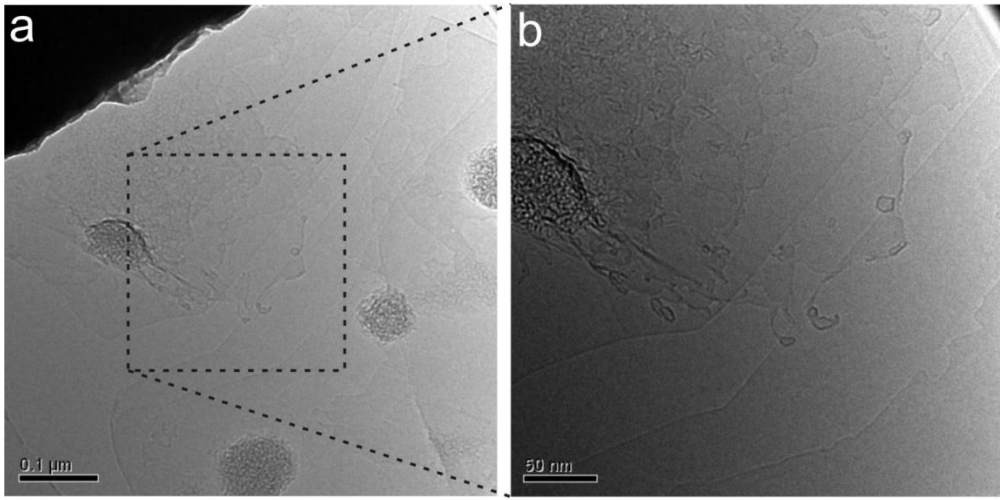

**Figure S8.** (a) Overview TEM image of a device that is electrically connected by two separated but overlapping graphene flakes. (b) Zoom in into the region of interest marked by the dashed square in panel (a).

**Section 8. Additional TEM images and current vs voltage data of the device in figure 8**

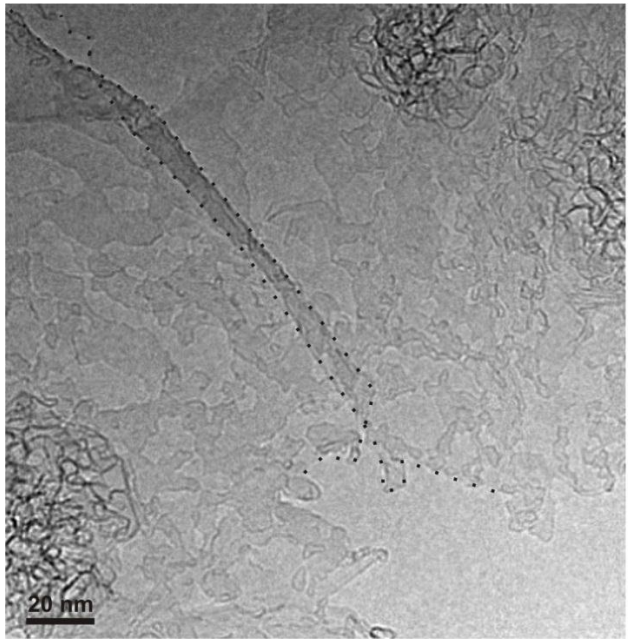

**Figure S9.** (a) Overview TEM image of the device in figure 8. The dotted lines are a guide to the eye of the borders of the two few-layer graphene samples and a region where the two flakes are overlapping.

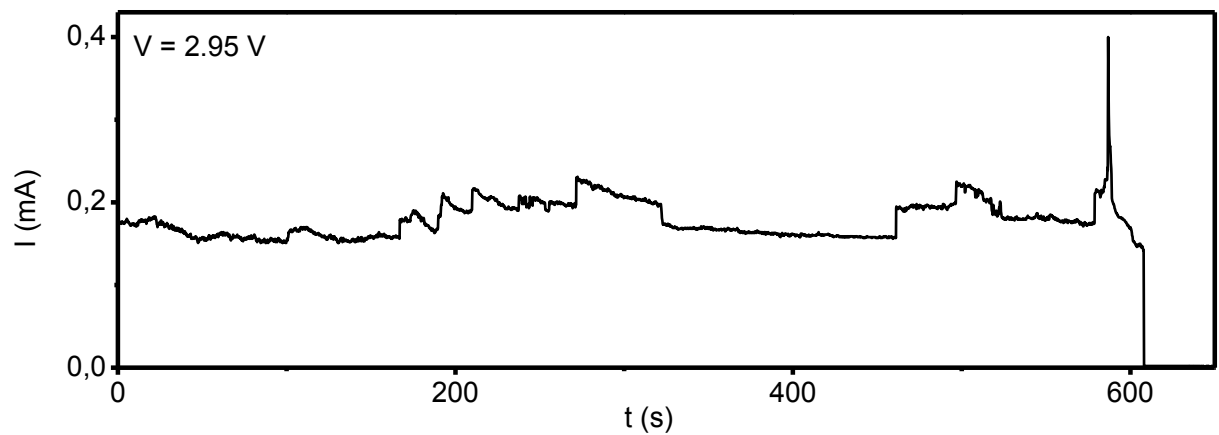

**Figure S10.** Current vs. voltage data showing the conductance changes that occur when the two overlapping regions in video S6 move relative to each other. Eventually the two pieces merge and heal into a seamless graphene sheet (see figure 8 of the manuscript, which corresponds to the last 38 seconds before rupture).

**Section 9. Multi-wall carbon nanotube formation at the edges of few-layer graphene after electrical breakdown**

Remarkably, at the edges of electro-burned suspended few-layer graphene we usually find MWNTs. The edges of the few-layer graphene already formed half a MWNT (nested BLEs, see main text). After electrical breakdown of graphene at high temperatures, the few layer graphene in the broken part rolls up completely to form a MWNT thus making the edges more stable.

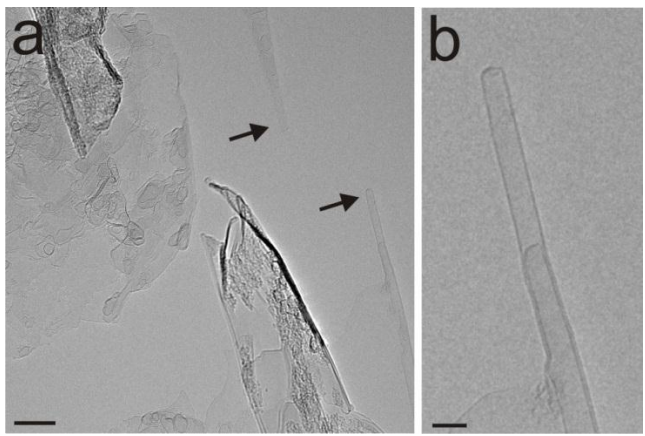

**Figure S11.** (a) TEM image of a graphene flake after electrical breakdown. The arrows point at nanotubes formed at the border of the flake. The scale bar is 20 nm. (b) The nanotube corresponds to the lower arrow in panel (a). The scale bar is 5 nm.

REFERENCES.